\documentclass{ifacconf}

\usepackage{graphicx}      
\usepackage{natbib}        
\usepackage{amsmath}
\usepackage{amssymb}
\usepackage{amsfonts}
\usepackage{mathrsfs}

\usepackage{algorithm}
\usepackage{xcolor}
\usepackage{tcolorbox}  
\usepackage{enumitem}
\usepackage{svg}
\usepackage{xpatch}
\usepackage{tikz}
\usepackage{comment}
\usepackage{svg}
\usepackage{tablefootnote}

\begin{document}
\begin{frontmatter}

\title{Least Costly Space-Filling Experiment Design for the Identification of a Nonlinear System\thanksref{footnoteinfo}} 

\thanks[footnoteinfo]{Funded by the European Union (ERC, COMPLETE, 101075836). Views and opinions expressed are however those of the author(s) only and do not necessarily reflect those of the European Union or the European Research Council Executive Agency. Neither the European Union nor the granting authority can be held responsible for them.}

\author[First]{Máté Kiss}  
\author[First]{Maarten Schoukens}
\author[First,Second]{Roland Tóth}

\address[First]{Control System group, Eindhoven University of Technology,
Eindhoven, the Netherlands}
\address[Second]{Systems and Control Laboratory, Institute for Computer Science
and Control, Budapest, Hungary}

\begin{abstract}
The quality of an estimated nonlinear model highly depends on the data that was used for its identification. By using a Gaussian Process-based optimal input design approach, a so-called space-filling dataset can be generated in the feature space of a preliminary model of the system. The design method is applicable for a broad type of signals and models and also incorporates information measures through optimality criteria into the signal design. However, the resulting input design can be costly to apply to the real system. The goal of this paper is to propose a space-filling input design that can minimize the experimentation cost in terms of a user defined measure, while still guaranteeing a prescribed level of space-fillingness. Through a Monte Carlo simulation study, we demonstrate that the proposed method can appropriately shape the excitation signal to significantly reduce the experimental cost while the identified model performance remains adequate.
\end{abstract}


\end{frontmatter}

\section{Introduction}
The goal of a system identification process is to estimate a mathematical model from noisy measurement data. Such a process strongly relies on the quality and informativity of the available data. While this is already an important issue for \emph{linear time-invariant} (LTI) system identification, it becomes even more important for nonlinear system identification. An LTI model may be seen as a hyperplane in the feature space, a nonlinear model is characterized by a manifold \citep{Schoukens19}, thus being more difficult to extrapolate. Therefore a nonlinear model is much more sensitive to modeling errors and assumptions on the model structure.

The leading classical input design strategy, known as optimal input design, aims to minimize the variance of the identified model parameters \citep{Bombois2021}. For this purpose an optimality criterion, often based on the Fisher information matrix, is defined. Common approaches are known as A-, D- and V-optimality \citep{atkinson1992}. However, existing design methods are mostly focused on LTI systems, using the coarse assumption of an unbiased estimator \citep{Bombois2021,annergren2017application}.
On the other hand, optimal input design for nonlinear system identification is much less understood due to the simultaneous dependence on time- and frequency-domain characteristics \citep{Cock_D_optimal}. Actual design approaches have been mostly studied in detail for simple nonlinear systems, such as the Hammerstein and Wiener classes \citep{Colin20,colin2020data,Cock_Phd,Forgione2014}.

However, as outlined previously, the leading challenge in black-box nonlinear system identification is not to have a small variance on the parameter estimates, but rather to ensure that the model obtains a good quality over the considered range of operation because the dominant source of errors is that the system is rarely part of the model class \citep{Schoukens19}. Therefore, there is always a discrepancy between the model and the system due to modeling approximations. A so called space-filling input, which creates a dataset that sufficiently covers the feature space of the model \citep{herkersdorf2025online,liu2025space,vater2024differentiable}, is capable to ensure that a model identified using this dataset behaves well over the full region of interest \citep{kiss2024space}.

Most recent space-filling input design methods are focusing purely on the coverage of the region of interest \citep{herkersdorf2025online,herkersdorf2024optimized,vater2024differentiable,kiss2024space}; nonetheless, an experiment entails an economical cost as well. The so-called least costly experiment design paradigm \citep{bombois2012design,bombois2004least} has proven for linear models that the experiment cost can be reduced while guaranteeing almost the same identified model performance compared to optimal design. Following that ideology, the present paper addresses the challenge of how to design a space-filling signal with the lowest cost such that a given level of space-fillingness is guaranteed.

The structure of the paper is the following: Section~\ref{sec:ProblemFormulation} formulates the proposed least costly and the classical space-filling design method. Afterward in Section~\ref{sec:SimulationStudy}, performance of the proposed method is demonstrated in a Monte Carlo simulation study and compared to a non-space filling and to a classically designed space-filling signal. Finally, Section~\ref{sec:Conclusion} describes the conclusions.

\section{Problem Formulation}\label{sec:ProblemFormulation}
\subsection{System and Signal Class}
Consider a controllable, deterministic system model whose arguments, referred to as features, are represented by $x(k)$. The system model approximates the process output $y(k)$ for any specific feature. It characterizes our prior knowledge of the system and represents the chosen model for which we will later on design the space filling input signal. We use the following input output representation as a system model:
\begin{equation}\label{eq:systemModel}
\begin{aligned}
    y(k) \!&=\! f(y(k-1), \ldots, y(k-m_\mathrm{y}), u(k-1), \ldots, u(k-m_\mathrm{u}))\\
          &= f(x(k)),
\end{aligned}
\end{equation}
where $u(k)\in\mathcal{U}\subseteq\mathbb{R}^{n_\mathrm{u}}$ and $y(k)\in\mathcal{Y}\subseteq\mathbb{R}^{n_\mathrm{y}}$ are the input and the output of the system at time instant $k \in \mathbb{I}_1^N$. We consider a single-input single-output case in the present work, thus $n_u\!=\!1$ and $n_y\!=\!1$. The maximum time lags for the respective signals are defined by $m_\mathrm{y}$, $m_\mathrm{u}$. The feature space of the model is defined by the vector $x(k) = \mathrm{vec}(u(k-1),...,u({k-m_\mathrm{u}}),y(k-1),...,y({k-m_\mathrm{y}}))\in\mathcal{X}\subseteq\mathbb{R}^{n_\mathrm{x}}$, where $n_\mathrm{x}=n_\mathrm{u}m_\mathrm{u}+n_\mathrm{y}m_\mathrm{y}$, and  $\mathcal{X}:\mathcal{U}\times\mathcal{Y}$ denotes the joint delayed input-output space. Consequently the dataset is defined as $\mathcal{D}_N\!\in\!\{ \mathcal{X}\!\times\!\mathcal{Y} \}^N$. The nonlinear function $f(\cdot):\mathcal{X}\!\to\!\mathcal{Y}$, $f\in\mathcal{C}_1$ represents the system model. As a remark, we highlight that the proposed input design approach is not limited to input output forms, also a nonlinear state-space representation can be considered \citep{liu2025space,kiss2024space}.

Space-filling input design aims to optimize a parameter vector $\theta\!\in\!{\Theta}$ w.r.t. an objective function that corresponds to an input sequence $u_{\theta}$ with a desired space-filling density. 
Let the parametrized input signal be given by:
\begin{equation}
   u_\theta: \Theta \to \mathbb{R} ,
   \label{eq:inputSignalDefinition}
\end{equation}
where $u_\theta\!\in\!\mathcal{C}^1$ w.r.t. the signal parameters $\theta\!\in\!{\Theta}$. The signal is bounded for all $k\!\in\!\mathbb{T}$ with the finite time index set $\mathbb{T} = \{ 1,\ldots, N\}$ and with the compact parameter set ${\Theta}\!\subseteq\!\mathbb{R}^{n_\mathrm{\theta}}$. Other than the differentiability and boundedness, there are no more restrictions on the input signals. In other words, (\ref{eq:inputSignalDefinition}) can represent a wide class of parametric input signals with a broad selection of signal parametrization. Some parametrization can be chosen to embed signal constraints (e.g., a multisine parametrization to embed spectral constraints).

\subsection{Definition of space-fillingness}
\begin{figure}[t]
    \centering
    \includegraphics[width=\columnwidth]{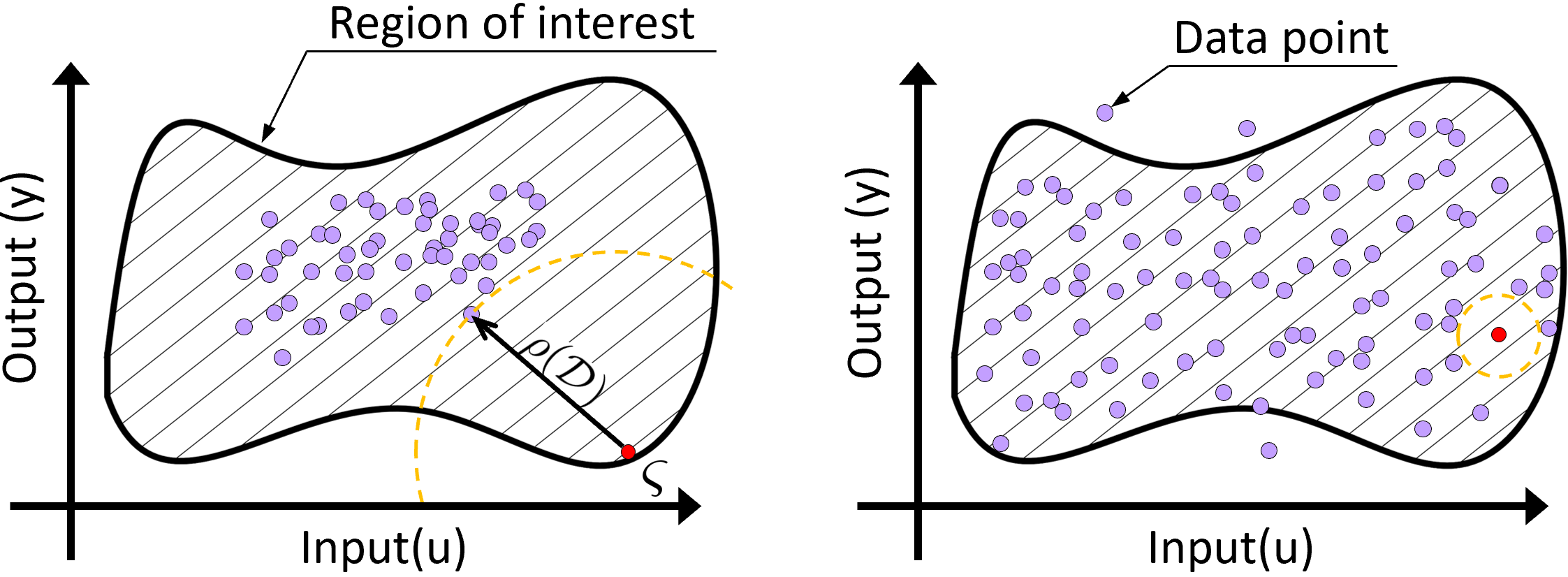}
    \caption{The concept of space-fillingness in the feature space: The striped domain denotes the region of interest ($\tilde{\mathcal{X}}$), purple dots denote the dataset ($\mathcal{D}$), while the yellow circle denotes the largest sphere between the data points with center $\varsigma$.}
    \label{fig:SpaceFillingness}
\end{figure}

The goal of a space-filling design is that any data point in the feature space is never too far from a design point ($\varsigma$) where we intend to evaluate our function \citep{pronzato2012design}. This means, if the model \eqref{eq:systemModel} is excited with an input signal $\{u(k)\}_{k=1}^{N}$, the generated data set $\mathcal{D}$ is scattered in the region of interest $\tilde{\mathcal{X}}$, where the region of interest is a compact subset of the feature space $\tilde{\mathcal{X}} \subseteq \mathcal{X}$. Choose a distance metric $d(\cdot,\cdot):\mathcal{X}\times\mathcal{X}\to\mathbb{R}_0^{+}$ and denote the minimum distance $d(\varsigma,\mathcal{D})=\min_{x_i\in\mathcal{D}}d(\varsigma,x_i)$ w.r.t. $\varsigma$. Then, we can define the so-called covering radius:
\begin{equation}\label{eq:CR}
\begin{gathered}
    \rho(\mathcal{D}) = \max_{\varsigma \in \tilde{\mathcal{X}}} d(\varsigma,\mathcal{D}),
\end{gathered}
\end{equation}
where $d(\varsigma,\mathcal{D})$ indicates the radius of the largest sphere,in terms of a given distance measure $d(\cdot,\cdot)$ (e.g. Euclidean distance), with a center $\varsigma$ in the region of interest such that no data point is contained in it (see left image in Fig.~\ref{fig:SpaceFillingness}).

A data set that minimizes $\rho(\mathcal{D})$ is sought such that the resulting dataset is well scattered inside the region of interest (see right image in Fig.~\ref{fig:SpaceFillingness}). If there exists a constant $\epsilon$\ such that $\rho({\mathcal{D}})<\epsilon, ~\epsilon>0$, then $\mathcal{D}$ is said to have at least $1/\epsilon$ density w.r.t $\tilde{\mathcal{X}}$.

\subsection{Preliminaries on Gaussian Process Regression}
Let $\mathcal{D}_N\!=\!\{X, Y \}$ be a dataset, where the matrix of features is $X = [x(1),\cdots,x(N)]^{\top}\!\in\!\mathbb{R}^{N \times n_x}$, with $x(k)\!=\![u(k-1),\ldots, u(k-{m_\mathrm{u}}), y(k-1),\ldots, y(k-{m_\mathrm{y}})]$, $k\!\in\!\mathbb{I}_1^N$. The scalar outputs are collected in $Y\!=\! [y(1),\cdots,y(N)]^{\top}\!\in\!\mathbb{R^{N}}$ generated by $y(k)\!=\!f(x(k)) + e(k)$, where $f:\mathbb{R}^{n_x} \to \mathbb{R}$ is an unknown function and $e(k) \sim \mathcal{N}(0,\sigma_{e}^2)$ is an i.i.d. gaussian noise.
The core idea of GP-based estimation of $f$ is to consider the candidate estimates $g$ belonging to a GP, seen as a prior distribution. Then using $\mathcal{D}_N$ and this prior, a predictive GP distribution of $g$ is computed that provides an estimate of $f$ in terms of its posterior mean and variance function.\\
A scalar-valued \emph{Gaussian Process} $\mathcal{GP} : \mathbb{R}^{n_x} \to \mathbb{R}$ assigns to every point $x \in \mathbb{R}^{n_x}$ a random variable $g(x)$, such that, for any finite set $\{x_i\}_{i=1}^N \in \mathbb{R}^{n_x}$, the joint probability distribution of $g(x_1), \cdots,g(x_N)$ is a Gaussian. Due to this property, $g \sim \mathcal{GP}(\mu,\kappa)$ is fully determined by its mean $\mu(x)$ and its covariance $\kappa(x,x')$ function, where $x,x' \in \mathbb{R}^{n_x}$.
We assume that the prior mean is zero $(\mu(x) = 0)$ and the covariance of the distribution can be well described by a \emph{squared exponential} (SE) kernel:
\begin{equation}\label{eq:kernel}
    \kappa\left({  {x}},  {  {x}}'\right)\!=\!\sigma_{\mathrm{f}}^{2} \exp \left(-\frac{1}{2}( {  {x}}\!-\!  {  {x}}')^{\top} {\Lambda}^{-1}( {  {x}}\!-\! {  {x}}')\right)\!, 
\end{equation}
where $\sigma_f^2 \in \mathbb{R}$ is a scaling factor and $\Lambda = \mathrm{diag}(d_1^2,...,d_{n_\mathrm{x}}^2)$ determines the smoothness of the kernel. Based on $\mathcal{D}_N$ and the prior $g \sim \mathcal{GP}(\mu,\kappa)$, the predictive distribution for $g(x_\ast)$ at a test point $x_\ast$ is the posterior $\mathbb{P} \left( g(x_{\ast}) \mid \mathcal{D}_N, x_{\ast} \right) = \mathcal{N}(\hat{\mu}(x_{\ast}),\hat{c}(x_{\ast}))$ characterized by
\begin{subequations}
    \begin{align}
        \label{eq:postMean} \hat{\mu}(x_\ast) &= \kappa(x_\ast,X) K_{N}^{-1} Y,\\
        \label{eq:postVar} \hat{c}(x_\ast) &= \kappa(x_\ast, x_\ast)
        - \kappa(x_\ast,X) K_{N}^{-1} \kappa(x_\ast,X)^{\top},
    \end{align}
\end{subequations}
where $[K_{N}]_{i,j} = \kappa(x_i, x_j) \in \mathbb{R}^{N \times N}, i \in \mathbb{I}_1^N, j \in \mathbb{I}_1^N$ is the Gram matrix.
Tuning of the kernel hyperparameters (i.e., $\sigma_f^2,\Lambda$) can be achieved with a wide range of methods listed in \citet{rasmussen2003gaussian}.
The next subsection continues with the introduction of the GP-based space-filling cost function.

\subsection{V-Optimal Space-Filling Cost Function} \label{sec:V-opt}
According to the work of \citet{liu2025space}, by considering a latent \emph{hypothetical} model $\hat{y}(k) = \hat{f}(x(k))$ for the system \eqref{eq:systemModel}, the information content of the experiment can be measured in terms of the uncertainty of this hypothetical model that would result as the posterior GP estimate based on the given data.
Upon applying the input sequence $\{ u_k \}_{k=1}^N$ on the assumed model of the system \eqref{eq:systemModel}, the dataset $\mathcal{D}_N$ can be obtained. From this dataset $\mathcal{D}_N$, we construct the matrix of features $X_N$. Given $X_N$, the resulting posterior behavior is given by $\hat{f}\!\sim\!\mathcal{GP}\left({\hat{\mu}}, \hat{c}\right)$ with the posterior covariance matrix.
Furthermore, to avoid the evaluation of the cost function over a continuous domain, the region of interest $\tilde{\mathcal{X}}$ is discretized by $M$ number of anchor points. Thus, the anchor dataset is defined by $\mathcal{D}_I = \{ \tilde{x}_i \}_{i=1}^M$ such that all elements in the set are distinct. Since the elements are distinct in $\mathcal{D}_I$, their distribution define the $1/\epsilon$ density in $\tilde{\mathcal{X}}$.

A space-filling promoting cost function is achieved by using a kernel $\kappa(x,x')$ that is monotonically decreasing w.r.t. $||x-x'||$ where $x$ is considered as a data point and $x'$ as an anchor point \citep{liu2025space}. Hence, the distance between $x$ and $\tilde{x}$ decays as the distance metric $||x-\tilde{x}||$ increases. By choosing \eqref{eq:kernel} as the kernel, the posterior covariance matrix of the \emph{hypothetical} model $\hat{f}$ is defined as
\begin{align}
   \hat{c}(\hat{f} \mid \tilde{x}_i, D_N) &= \kappa(\tilde{x}_i, \tilde{x}_i) \notag \\
   &-\! \kappa(\tilde{x}_i, X) K_{N}^{-1} \kappa(\tilde{x}_i, X)^\top .
\end{align}
Choosing the information metric of the experiment design to be the average posterior variance  of the model prediction $\hat{f}$ over the design domain $\tilde{X}$, translates to the so called V-optimality criterion \citep{rainforth2024modern,pronzato2013design,atkinson1992}. It is expressed as the following scalar valued space-filling cost function, namely the average posterior variance evaluated at the anchor points:
\begin{equation}\label{eq:costFunction}
    \mathcal{V}(\theta; \mathcal{D}_N(\theta)):=\frac{1}{M} \sum_{i=1}^{M} \hat{c}(\hat{f} \mid \tilde{x}_i,\mathcal{D}_N(\theta)).
\end{equation}
Now, the space-filling input design problem can be expressed as an optimization problem to find a parameter vector $\theta$ that yields an input sequence $u_{\theta}(k)$ capable of generating a space-filling dataset $\mathcal{D}_N(\theta)$ by applying it to system \eqref{eq:systemModel} with the given initial condition $x_\circ$. Such that, the resulting cost \eqref{eq:costFunction} is minimal, ensuring the space-filling behavior inside the region of interest $\tilde{X}$:
\begin{subequations}\label{eq:optimizationProblem}
\begin{align}
\underset{\theta\in{\Theta}}{\text{min}} 
 & \quad \mathcal{V}({\theta};\mathcal{D}_N(\theta)) \label{eq:opt1}\\
 \text{s.t.} & \quad x(0)=x_\circ,\\
 & \quad y(k) \!=\! f(x(k)), \quad k\!\in\!\{0,\cdots, N \}\\
 &\quad \mathcal{D}_N(\theta)\!=\!\{x(k),y(k)\}_{k=1}^{N}.
\end{align}
\end{subequations}
While standard space-filling designs aim to minimize the distance metric \eqref{eq:CR}, the GP-based approach makes use of the fact that the covariance function inherently defines a distance metric based on the data distribution. In this minimax-type of space-filling design, the decision variable $\theta$ is optimized using a space-filling cost $\mathcal{V}(\mathcal{D}_N)$ , which incorporates the covariance function and, in turn, influences $\mathcal{D}_N$ to optimize the space-filling measure $\rho(\mathcal{D}_N)$. For the more interested readers on the GP-based space-filling input design we refer to the work of \citet{liu2025space,kiss2024space}. Since the proposed input design approach also minimizes a metric derived from the variance of the estimate, we refer to it in this paper as \emph{classical} space-filling input design by using the index $Cl$.

\subsection{Least Costly Space-Filling Input Design Problem}\label{sec:LC}
In this section, we will formulate the space-filling input design problem presented in Sec.~\ref{sec:V-opt} into a least costly optimization problem. It has been shown in the work of \citet{bombois2012design} that, at least for linear systems, inclusion of the experiment cost in the design problem can lead to a significantly lower experimental cost without compromising significantly on the input quality or on the quality of the estimated models using the least costly input. By pursuing the same ideology, the cost of the space-filling input design
\begin{equation}\label{eq:costDefinition}
    \mathscr{C}(\theta):\Theta \to \mathbb{R} \quad \mathrm{where} \hspace{1mm} \mathscr{C}\!\in\!\mathcal{C}_1,
\end{equation}
can be expressed as a function of the decision variable $\theta$. Possible choices for the measure of $\mathscr{C}(\theta)$ may be the magnitude of the input signal or its power, defined as:
\begin{equation}\label{eq:costPower}
    \mathcal{P}_u(\theta) = \frac{1}{N} \sum_{k=1}^N u_{\theta}^2(k) .
\end{equation}
To ensure a specific level of coverage within the region of interest, the space-filling accuracy constraint has to be defined as a function of the decision variable $\theta$. As demonstrated in the work of \citet{liu2025space}, the cost function \eqref{eq:costFunction} is linked to the space-fillingness; therefore, in this work, it is adopted as the accuracy constraint.

In this context, the least costly space-filling input design aims to find a parameter vector $\theta$ that results in the input sequence $u_\theta^\mathrm{LC}(k)$ with the lowest experiment cost $\mathscr{C}(\theta)$ while guaranteeing a specific level of space-fillingness $\gamma$. Consequently, the following optimization problem arises:
\begin{subequations}
\begin{align}
\underset{\theta\in{\Theta}}{\text{min}} & \quad \mathscr{C}(\theta) \label{eq:opt3}\\
 \text{s.t.} & \quad \mathcal{V}({\theta};\mathcal{D}_N(\theta)) \leq \gamma ,
\end{align}
\end{subequations}
where the space covering is expressed by the V-optimal cost function \eqref{eq:costFunction} and $\gamma$ is a user chosen threshold for that. The design process starts with a feasibility problem:
\begin{equation}\label{eq:feasability}
    \mathcal{V}({\theta};\mathcal{D}_N(\theta)) < \gamma ,
\end{equation}
to find a $\theta_\circ\!\in\!\Theta$ such that \eqref{eq:feasability} is satisfied. 
A feasible starting point is achieved by solving \eqref{eq:optimizationProblem} which yields an input signal $u_{\theta_\circ}$. Applying this signal to system \eqref{eq:systemModel} returns a dataset from which a $\gamma_{\circ}$ can be computed using \eqref{eq:costFunction}.
Then, $\gamma$ is chosen to be larger than $\gamma_{\circ}$ by a user-defined margin. The chosen $\gamma$ value has to give sufficient room to the least costly design to make a trade-off between space-fillingness and the cost of the experiment; e.g., we used a 5\% margin in the simulation study below.

\section{Simulation Study}\label{sec:SimulationStudy}
The proposed least costly input design approach is tested in a Monte Carlo study with 50 realizations, where we compare it to the classical\footnote{\label{fn:classicalTerm}The approach is termed \emph{classical} because its optimization objective is derived from a variance/covariance matrix of the model estimates (see Sec.~\ref{sec:V-opt}) \citep{bombois2012design}.} space-filling approach. In both the least costly and in the classical design, the goal is to identify accurate models such that the input signal power is minimal. First, the quality of designs is compared in Sec.~\ref{sec:ExperimentDesign}. Afterward, their effectiveness is tested in Sec.~\ref{sec:Identification} by identifying nonlinear output error (NOE) models based on the resulting datasets.

\subsection{The Considered System}
\begin{figure}[t]
    \centering
    \includegraphics[width=3.5cm]{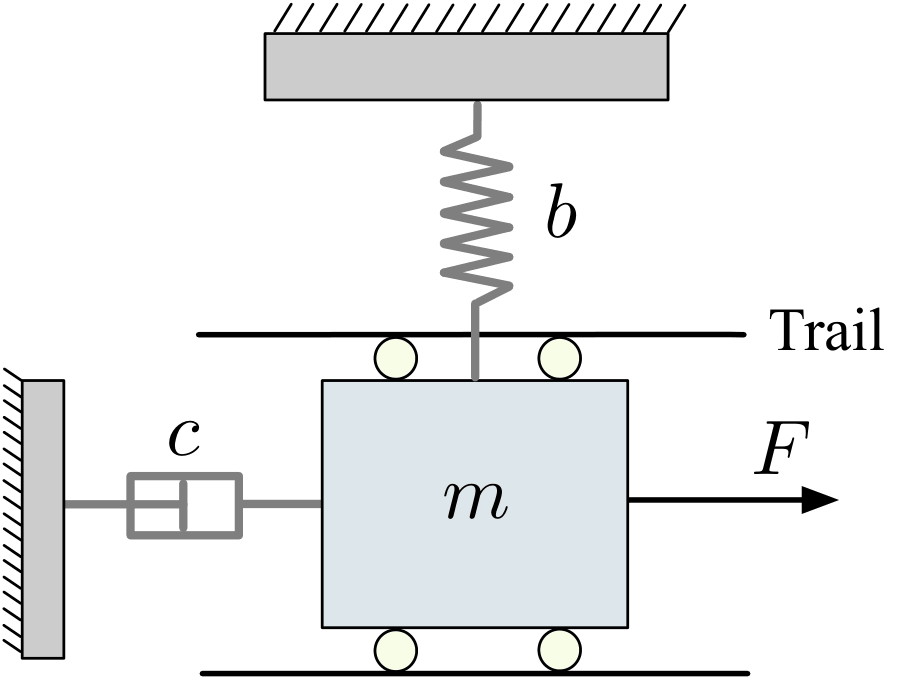}
    \caption{Nonlinear mass-spring-damper system.}
    \label{fig:MSD}
\end{figure}

The MSD system (Fig.~\ref{fig:MSD}) is considered in the following form:
\begin{equation} \label{eq:MSD}
\dot{x}_1\!=\! x_2,
\dot{x}_2 \!=\! \!\frac{1}{m}\!\left(\!F\! - \!s\frac{x_1}{\sqrt{x_1^2+a^2}} \left(\!\sqrt{x_1^2\!+\!a^2} \!-\! l\!\right) \!-\! c x_2\!\right)
\end{equation}
where $x_1$ and $x_2$ denote the position and velocity. The compressed and tensionless lengths of the spring are represented by $l\!=\!0.17$~m and $a\!=\!0.25$~m. The mass is given by $m\!=\!5$~kg, while $s\!=\!800$~N/m denotes the spring stiffness and $c\!=\!10$~Ns/m denotes the damping coefficient. $F$ is the excitation signal of the system described in Sec.~\ref{sec:InputSignal}. The state-space model is transformed into a discrete-time input-output representation $y(k)=f(y(k-1),dy(k-1),u(k-1))$, where $y(k)$ is the displacement (output of the model) and 
 $dy(k)\!=\!y(k)-y(k-1)$ is the output increment, from which the velocity can be calculated.

\subsection{Input Signal and Region of Interest}\label{sec:InputSignal}
In this example we use a multisine signal $u_\mathrm\!\in\!u_{\Theta}$ with parameters $\theta\!=\!\mathrm{vec}(\{\mathrm{A}_l,\varphi_l\}_{l=l_\mathrm{min}}^{l_\mathrm{max}})$: 
\begin{equation}\label{eq:multisine}
    u(k; \theta) \!=\! \sum_{l=l_\mathrm{min}}^{l_\mathrm{max}} \mathrm{A}_l \sin\left(2\pi l \frac{f_0}{f_s}k + \varphi_l\right).
\end{equation}
Between the range $f_\mathrm{{min}}$=1~Hz and $f_\mathrm{{max}}$=10~Hz, every $\mathrm{7}^{th}$ frequency line is excited which corresponds to $L\!=\!14$ in total with $l_\mathrm{min}\!=\!12$, $l_\mathrm{max}\!=\!103$. The phases are initialized using a uniform random distribution $[0,2 \pi[$, while $f_0\!=\!f_s/N$ corresponds to the frequency resolution with the sampling frequency $f_s$=100~Hz and $N$=1024 data samples per period. 
The amplitudes $\{\mathrm{A}_l\}_{l=l_\mathrm{min}}^{l_\mathrm{max}}$ are parametrized such that  their value can change per sample point. In our simulation study, every excited frequency starts with the initial amplitude of 8~N.

The region of interest is a rectangle in the 2-dimensional delayed output space $\mathcal{\tilde{X}} \!=\! \{ (y,dy) \!\in\! \mathbb{R}^2 \mid y \!\in\! [-0.1,0.1], \; dy \!\in\! [-0.8,0.8] \}$ (see Fig~\ref{fig:ExperimentDesign}(e) for an example). It is represented by 5 equally distanced anchor points $\tilde{x}_i$ along each dimension of the space, giving $M=25$ anchor points in total.
For the considered MSD system in \eqref{eq:MSD}, the input signal enters the motion equations linearly. Therefore, one may omit the input signal dimension from the region of interest so the space-filling input signal is designed only for the nonlinearities. In addition, this choice also allows for better visualization.
The kernel widths are chosen to be equal with the adjacent anchor point distance in the corresponding dimension $\Lambda = \mathrm{diag}(0.05,0.40)$, the scaling factor is $\sigma_{\mathrm{f}}^{2} = \sqrt{10}$ and $\sigma^2_{\epsilon}=1$.

\subsection{Input Design}\label{sec:ExperimentDesign}
\begin{figure}[t]
    \centering
    \includegraphics[width=\columnwidth]{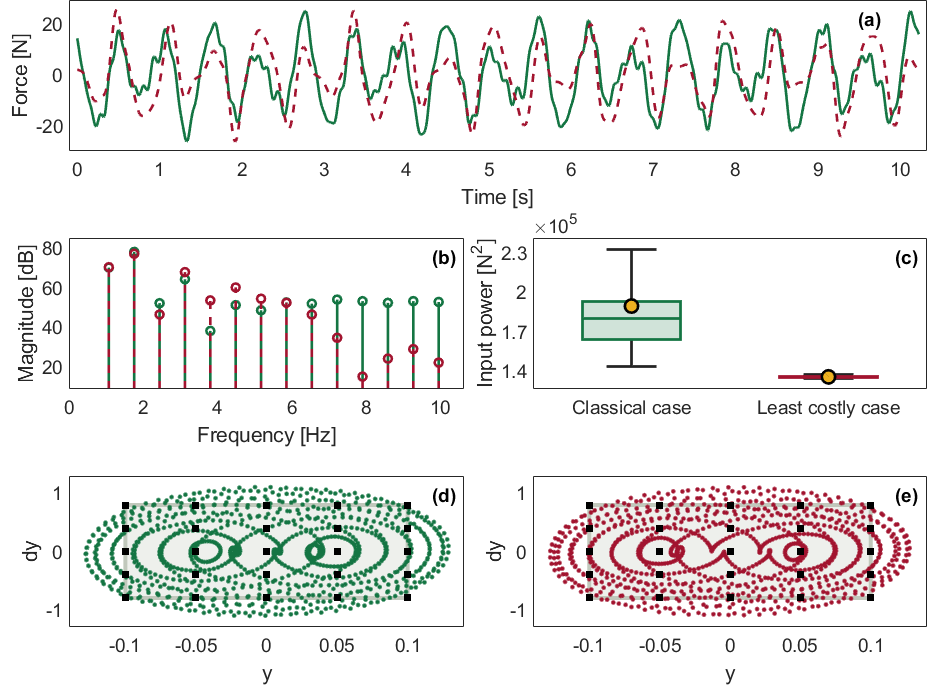}
    \caption{(a) Least costly $u^{LC}_{\theta}$ (red) and classical $u_{\theta}^{Cl}$ (green) space-filling signal, (b) frequency plot of $u^{Cl}_{\theta}$ (green) and $u^{LC}_{\theta}$ (red), (c) box plot of input powers with their mean value (yellow circle), (d,e) classical $\mathcal{D}^{Cl}$ and least costly $\mathcal{D}^{LC}$ (red) space-filling dataset.}
    \label{fig:ExperimentDesign}
\end{figure}
The space-filling signal is designed based on a linear approximation of the nonlinear mass-spring-damper (MSD) system \eqref{eq:MSD}. While our approach also can handle nonlinear models, using a linear approximate model illustrates that the method keeps on performing well for this example, even in the absence of full system knowledge. This linear model is then used as the system model \eqref{eq:systemModel} to create a space-filling signal, first using the least costly input design method based on Sec.~\ref{sec:LC} and second using the classical approach based on Sec.~\ref{sec:V-opt}. 
Every Monte Carlo realization starts with a newly initialized multisine and with a new linear approximation of the nonlinear MSD. The phases are drawn from the same uniform distributions and the amplitudes are initialized to 8~N. In each realization, the least costly design is obtained by first solving the feasibility problem \eqref{eq:feasability} that returns a $\gamma_\circ$ value. Then, a new value of $\gamma$ is set for the current realization to be 5\% larger than $\gamma_\circ$. The optimization is accomplished with the Matlab function \verb+fmincon+ using numerically computed gradients.

Figure~\ref{fig:ExperimentDesign} shows one example for the least costly $u^{LC}_{\theta}$ and one for the classical signal $u^{Cl}_{\theta}$ alongside the obtained space-filling data sets $\mathcal{D}^{LC}$ and $\mathcal{D}^{Cl}$. As depicted in Fig.~\ref{fig:ExperimentDesign}(c) the mean input power of the classical signals, $\mathcal{P}_{Cl,\mathrm{mean}}\!=\!1.89\!\times\!10^{5}~\!N^2$, is 40\!~\!\% larger than in case of the least costly signals $\mathcal{P}_{LC,\mathrm{mean}}\!=\!1.35\!\times\!10^{5}~\!N^2$. However, a near equivalent space-filling behavior can be seen in Fig.~\ref{fig:ExperimentDesign}(d) and (e), which in average over the 50 realizations is equal to $\mathcal{V}_{Cl,\mathrm{mean}}\!=\!0.0428$ and $\mathcal{V}_{LC,\mathrm{mean}}\!=\!0.0429$. Since the difference is less than 1\!~\!\%, the signals can be considered to be equivalently good in terms of information content. Concerning the covering radius the same conclusion can be deduced, the difference is 3\!~\!\% with $\rho_{\mathrm{Cl,mean}}\!=\!0.030$ and $\rho_{\mathrm{LC,mean}}\!=\!0.029$. One could expect a significantly higher identified model quality for signals with more power. As the next section will show in the examples, this is not obviously the case.

\subsection{System identification}\label{sec:Identification}
For each Monte-Carlo realization, there are 3 models identified: one with the least costly space-filling dataset $\mathcal{D}^{\mathrm{LC}}$, one with the classical space-filling dataset $\mathcal{D}^{\mathrm{Cl}}$ and one with the initial dataset $\mathcal{D}^{\mathrm{Ini}}$ generated by the unoptimized random phase multisine signal. The nonlinear output error (NOE) model class is considered during identification:
\begin{subequations}
    \begin{align}
        y_0(k)\!&=\!f_{\eta}(y(k-1),\!\ldots\!,\!y(k-m_\mathrm{y})\!,\!u(k-1)\!,\!\ldots\!,\!u(k-m_\mathrm{u})) \\
    y(k)\!&=\!y_0(k)\!+\!e(k).
    \end{align}
\end{subequations}
The function $f_\eta$ is parametrized by a 1 hidden layer feedforward neural network with 8 neurons using a sigmoid activation function $\sigma(\cdot)$ such that:
\begin{equation}
    f_{\eta}(x(k)) = W_x \sigma \left( W_{fx} x(k)  + b_f \right) + b_x,
\end{equation}
with  $x(k) = \mathrm{vec}(y(k-1),\!\ldots\!,\!y(k-m_\mathrm{y})\!,\!u(k-1)\!,\!\ldots\!,\!u(k-m_\mathrm{u}))$. The model parameters are contained in $\eta$; i.e. $W_x, W_{fx}$ network weights and $b_f,b_x$ network biases which are estimated through the minimization of the mean squared error:
\begin{subequations}
\begin{align}
\label{eq:NOEcost} V_N(\eta) &= \frac{1}{N} \sum_{k=1}^{N} (y(k) - \hat{y}(k|\eta))^2, \\
\hat{\eta} &= \arg \min_{\eta} V_N(\eta),
\end{align}
\end{subequations}
where $\hat{y}(k|\eta)$ is the simulated output of the neural network given the parameter vector $\eta$, the system output $y(k)$ and $N$ total number of data samples. The cost function \eqref{eq:NOEcost} is minimized with a \emph{Levenberg-Marquardt} algorithm. As a measure of model quality, the simulation root mean square error (RMSE) is used to compare the results:
\begin{equation}
    e_{\text{RMSE}}\!=\!\sqrt{\frac{1}{N} \sum_{k=1}^{N} (y(k)\!-\!\hat{y}(k|\eta))^2},
\end{equation}
where $y(k)$ is the observed noise free test output and $\hat{y}(k\!\mid\!\eta)$ is the simulated output using the estimated model.

\begin{figure}[t]
    \centering
    \includegraphics[width=\columnwidth]{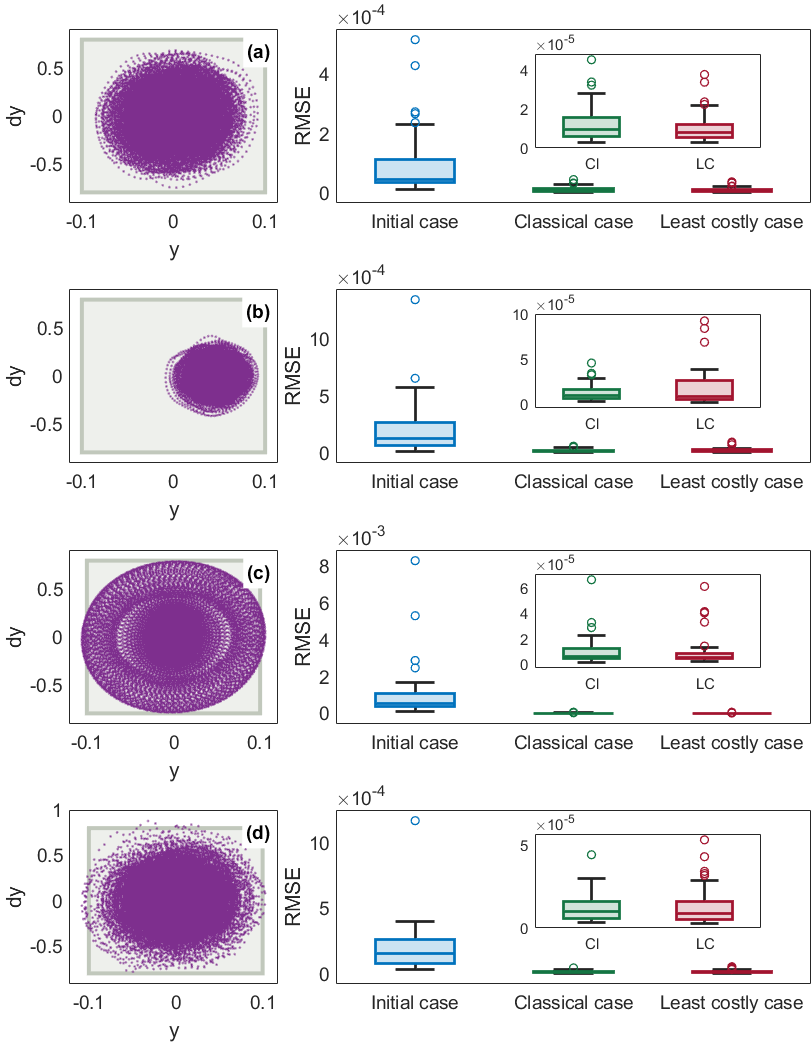}
    \caption{Results for different test sets: (a) multisine signal with constant 8~N amplitude, (b) multisine signal with constant 4~N amplitude, (c) swept signal with logarithmic amplitude, (d) white noise signal  and the corresponding model error boxplots on their right where the green shows the classical and the red shows the least costly RMS errors. The gray area denotes the region of interest $\tilde{\mathcal{X}}$.}
    \label{fig:IdentificationResults}
\end{figure}

To validate our results we use different types of test sets with $N_\mathrm{test}\!=\!2^{16}$ number of data samples per period and we excite all frequencies between $f_{\mathrm{min}}$ and $f_{\mathrm{max}}$ (see Fig.~\ref{fig:IdentificationResults}). The first test set is generated by a random phase multisine having the same signal parametrization and attributes that the initial multisine (i.e. meaning that the test signal is coming from the same input class) (Fig.~\ref{fig:IdentificationResults}(a)). Accordingly, the box plots of Fig.~\ref{fig:IdentificationResults}(a) illustrate the corresponding model errors coming from the 50 realizations. The following test set is also a multisine with the same signal parametrization that the space-filling signals, but with an amplitude of 4~N such that it discovers only a part of the region of interest (Fig.~\ref{fig:IdentificationResults}(b)). Figure~\ref{fig:IdentificationResults}(c) presents the dataset generated by a sweep signal with logarithmic amplitudes $u(k)\!=\!\sin \left(2\pi f_{\mathrm{min}} L \exp \left( k/L \right) \right)$ where $L\!=\!T/\ln\left(f_{\mathrm{max}}/f_{\mathrm{min}}\right)$ and $T$ denotes the duration of the signal. The last test set is generated by a white excitation with variance $\sigma_w^2\!=\!19$ of a uniform distribution (Fig.~\ref{fig:IdentificationResults}(d)).

Table~\ref{t:IdentificationResults} gives a summarizing overview of the median model errors coming from the 50 realizations. It can be seen that the resulting model errors from the least costly data sets are very similar to the classical one. In this particular study, we observe an average gain of 11\% in terms of model quality over the 4 test cases by using 40\% less power in our signals while the space-covering behavior remained the same as Fig.~\ref{fig:ExperimentDesign}(d) and (e) shows\footnote{\label{fn:lowerRMSEwithLC}We note that obtaining lower modeling errors with the least costly design than with the classical one is not guaranteed. However, the least costly design is optimized over more iterations, which may allow it to approach the global optimum more closely. Moreover, since the design model is linear, the optimum obtained w.r.t. this model does not necessarily correspond to the optimal design for the true system (see Example 1 of \citet{bombois2012design}).}.
The benefit of the space-covering input design can be well observed by comparing the results of the initial case to both the classical and least costly cases. The classical experiment design based estimated models have a 24.74 times and the least costly experiment design based estimated models have a 27.63 times lower model error in average over the four test cases. Due to the space-filling nature, the model quality remains similar for the 4 test cases. While the model quality varies a noteworthy amount over the four test cases using only the random phase multisine data $\mathcal{D}^{\mathrm{Ini}}$.

\begin{table}[t]
\centering
\caption{Median RMSE on the scale of $10^{-5}$ of the initial, classical and least costly cases.}
\begin{tabular}{|c||c|c|c|}
\hline
\text{Case} & \rule{0pt}{2.6ex}$e^{\mathrm{ini}}$ & $e^{Cl}$ & $e^{\mathrm{LC}}$ \\ \hline\hline
Multisine (8~N) & 4.73 & 0.93 & 0.80 \\ \hline
$\mathrm{Multisine_{shifted}}$ (4~N) & 12.58 & 0.93 & 0.88 \\ \hline
$\mathrm{Sweep_{log}}$ (6~N) & 52.84 & 0.61 & 0.53 \\ \hline
White noise & 15.23 & 0.98 & 0.88 \\ \hline
\end{tabular}
\label{t:IdentificationResults}
\end{table}

\section{Conclusion}\label{sec:Conclusion}
In this paper, a least costly space-filling input design approach has been proposed to create a signal with the minimal cost such that it still guarantees a certain level of coverage over the region of interest. Due to the rich input parametrization capabilities of the design approach, we could effectively inject power to regions of the system where required while limiting elsewhere. By including the experiment cost in the space-filling design problem, the experimental cost could be reduced significantly without compromising the input quality or the quality of the estimated models. Moreover, the space-filling characteristics of the dataset were preserved. As a result, the model identified using this dataset behaves well over the region of interest.
Compared to the state-of-the-art space-filling input design methods, the presented algorithm can provide a flexible choice for the experiment cost and for the signal parametrization as well without being limited to any specific signal class.

\bibliography{ifacconf}             

\end{document}